**Photoinduced Domain Pattern Transformation in Ferroelectric/Dielectric Superlattices**

Youngjun Ahn,[1] Joonkyu Park,[1] Anastasios Pateras, [1] Matthew B. Rich, [1] Qingteng Zhang,[1,*] Pice Chen,[1,*] Mohammed H. Yusuf,[2] Haidan Wen,[3] Matthew Dawber,[2] and Paul G. Evans[1,†]

[1] *Department of Materials Science and Engineering, University of Wisconsin-Madison, Madison, Wisconsin 53706, USA*
[2] *Department of Physics and Astronomy, Stony Brook University, Stony Brook, New York 11794, USA*
[3] *Advanced Photon Source, Argonne National Laboratory, Argonne, Illinois 60439, USA*

[*] Present address: Advanced Photon Source, Argonne National Laboratory, Argonne, Illinois 60439, USA.
[†] Electronic mail: pgevans@wisc.edu

The nanodomain pattern in ferroelectric/dielectric superlattices transforms to a uniform polarization state under above-bandgap optical excitation. X-ray scattering reveals a disappearance of domain diffuse scattering and an expansion of the lattice. The reappearance of the domain pattern occurs over a period of seconds at room temperature, suggesting a transformation mechanism in which charge carriers in long-lived trap states screen the depolarization field. A Landau-Ginzburg-Devonshire model predicts changes in lattice parameter and a critical carrier concentration for the transformation.

The formation and geometric pattern of nanodomains in ultrathin ferroelectrics depend on a sensitive balance of competing energetic contributions. Distinct domain morphologies result from the minimization of the free energy, which includes contributions from the depolarization field, electrical polarization, elastic energy, and strain gradients. Thermodynamic models based on Landau-Ginzburg-Devonshire (LGD) theory can be used to evaluate the stability of the system and to discover favorable configurations [1-3]. Among experimental realizations of ferroelectric nanodomains, superlattice heterostructures consisting of alternating ferroelectric and dielectric layers exhibit domain configurations and electrical properties that can be tuned by adjusting the layer composition, periodicity, and strain [4-8]. The key physical parameter of ferroelectric/dielectric superlattice heterostructures is the difference in the polarization of the ferroelectric and dielectric layers, which leads to the generation of a depolarization field. Mechanisms for tuning and screening the depolarization field have received significant attention [9-11]. The depolarization field of ultrathin layers can be screened by chemical adsorbates [12,13], charged oxygen vacancies [14], or metallic electrodes [15,16], resulting in changes in both the domain pattern and the atomic structure. Applied electric fields can similarly affect the domain pattern, including by introducing a transformation to a uniform domain configuration [10, 17].

The optical excitation of ferroelectrics results in a range of structural effects. Illumination can induce domain-wall motion in bulk ferroelectrics [18] or the production of a photovoltaic current [19]. The stress arising from optical absorption in metallic component of ferroelectric/metallic oxide superlattices can result in a complex time-dependent evolution of the polarization [20]. Phenomena induced by above-bandgap illumination of ferroelectric thin films include an expansion of the lattice following intense above-bandgap excitation [21-26]. The experimentally observed lattice expansion is linked to the large photoexcited charge carrier density, and exhibits a relaxation time approximately equal to the decay time of electron-hole pairs [24]. Optically induced effects can be coupled into other components of heterostructures, including at magnetic metal/multiferroic interfaces [27]. Mechanisms suggested for the expansion include the screening of the depolarization field by the migration of photoexcited charges to interfaces [24] and more localized charge carrier separation [25]. How the longer-range nanoscale organization of the polarization into domains responds to the optical illumination, however, has not yet been resolved. In this Letter, we report the discovery and physical mechanism of an optically induced transformation from a nanodomain configuration to a uniform polarization state in a $PbTiO_3/SrTiO_3$ superlattice (PTO/STO SL). Key aspects of the origin and nanoscale mechanism of the domain transformation are revealed by examining its dependence on the absorbed optical intensity and the dynamics of the re-establishment of the domain pattern.

The equilibrium room-temperature 180° stripe nanodomain pattern of a PTO/STO SL is illustrated in diagram (i) of Fig. 1(a). The PTO/STO SL system has well-defined ferroelectric properties, including low leakage and a systematic scaling of the Curie temperature and domain period with layer thickness and average composition [8,28]. The diagram in Fig. 1(a) includes only one direction of the in-plane domain periodicity. It is important to distinguish between the uniform polarization state reached by the optically induced transformation and the paraelectric phase observed above the Curie temperature $T_C$. As we show below, the optically induced uniform polarization state exhibits a lattice expansion (diagram (ii) of Fig. 1(a)), while the high-temperature paraelectric phase reached by heating without optical excitation results from the tetragonal-to-cubic transition at $T_C$ (diagram (iii) of Fig. 1(a)). Changes in the structure and domain configuration can be distinguished using x-ray diffraction. Figure 1(b) shows schematics of reciprocal space for (i) the nanodomain configuration, (ii) the optically induced uniform polarization state, and (iii) the paraelectric state above $T_C$. In the nanodomain configuration, nanodomains produce a ring of x-ray diffuse scattering in the $Q_x$-$Q_y$ plane around each Bragg reflection of the SL. The optically induced uniform polarization state has two key signatures: the domain diffuse scattering ring disappears and the SL Bragg reflection shifts to a lower $Q_z$. The reciprocal space map of the high-temperature paraelectric phase, in comparison, exhibits a contraction shifting of the SL Bragg reflection to higher $Q_z$.

The experimental geometry for the synchrotron x-ray diffraction study of the optically excited PTO/STO SL appears in Fig. 1(c). The heteroepitaxial PTO/STO SL consisted of a repeating unit of 8 unit cells of PTO and 3 unit cells of STO, and had a total thickness $h$ = 100 nm. The SL was deposited on a $SrRuO_3$ (SRO) bottom electrode on an (001)-oriented STO substrate using off-axis radiofrequency magnetron sputtering [8]. A PTO layer was deposited at the SL/SRO interface and the top layer of the SL was composed of STO. The effective relative dielectric constant and resistivity of the $PbTiO_3/SrTiO_3$ SLs were 464 and $1.5 \times 10^8$ Ω cm, respectively, measured at a frequency of 1 kHz and 10 mV rms excitation voltage using a capacitance bridge (Andeen Hagerling 2500A). The dielectric loss factor tan $\delta$ was 0.03. X-ray microdiffraction measurements were performed at station 7ID-C of the Advanced Photon Source [17], using a photon energy of 11 keV a focal spot with 355 nm FWHM using a Fresnel zone plate. The diffracted x-ray intensity was measured using a pixel-array detector (Pilatus 100K, Dectris Ltd.).

The optical excitation consisted of pulses at a wavelength of 355 nm, photon energy 3.5 eV, with 10 ps pulse duration and a repetition rate of 54 kHz. The illumination was at higher energy than the nominal optical band gaps of PTO and STO, 3.4 eV [29] and 3.2 eV [30], respectively. Since the recovery time as shown below is longer than the interval between pump pulses, the optical excitation can be regarded as a quasi-continuous and the results reported here are given in terms of time-average intensity. Optical pulses were transported to the sample stage using a multi-mode optical fiber and focused with an ultraviolet objective lens to allow spatial overlap with the x-ray beam [31]. The optical focus had approximately a Gaussian spatial profile with full-width-at-half maximum (FWHM) diameter of 110 μm.

Values of the absorbed optical intensity $I_{abs}$ were calculated using $I_{abs} = I_{in}(1-R)[1-\exp(-\alpha h)]$, where $I_{in}$ is the incident optical intensity, $R$ is the normal-incidence optical reflectivity of the SL, $\alpha$ is the effective optical absorption coefficient of the SL at 355 nm, and $h$ is the total SL thickness. The nominal incident optical intensity $I_{in}$ is obtained by dividing the total incident optical power by the FWHM area of the optical spot. The optical constants of the superlattice were estimated using the effective medium approximation from the refractive indexes of PTO and STO [32-34], giving a complex index of refraction of 3.32 + 0.15i for the superlattice. The computed reflectivity, absorption coefficient, and effective value of $I_{abs}/I_{in}$ were $R$ = 0.27, $1/\alpha$ = 194 nm, and $I_{abs}/I_{in}$=0.28, respectively. We neglected a small additional contribution due to reflectance at the SL/SRO interface, approximately 3% of the total absorbed intensity based on the optical constants of SRO [35].

A spatial map of the integrated intensity of the domain diffuse indicates that the optically induced disappearance of the domains is confined to the illuminated area. The spatial extent of the domain transformation is apparent in Figs. 2(a) and 2(b), which show the same area of the SL in the dark and during illumination with an absorbed average intensity of 1.3 $W/cm^2$. Optical excitation at this intensity leads to a reduction of the domain diffuse scattering by 64% at the center of the illuminated region.

Three-dimensional maps of reciprocal space were constructed by acquiring diffraction patterns at a series of x-ray incident angles and converting to reciprocal space coordinates. Figure 2(c) shows the x-ray intensity distribution in sections of reciprocal space at $Q_x$ = 0 (top) and at $Q_z$ = 3.128 Å$^{-1}$ (bottom) in the state without illumination. Domain diffuse scattering appears in the $Q_x$-$Q_y$ plane at a reciprocal-space distance from the SL structural reflections given by $\Delta Q_{xy} = 2\pi/\Lambda$, where $\Lambda$ is the domain period [36]. The center of mass of the domain reflections is at $\Delta Q_{xy}$ = 0.076 Å$^{-1}$, giving $\Lambda$ = 8.3 nm. Figure 2(d) shows a reciprocal space map near the (002) SL reflection acquired with optical intensity 1.3W/cm$^2$. This intensity is near the threshold for the optically induced transformation to a uniform polarization. During illumination, the SL reflection splits to lower $Q_z$ and the intensity of the domain diffuse scattering intensity decreases by 61%.

The illumination shifts the out-of-plane wavevector of the SL Bragg reflections to lower $Q_z$ and a decrease in the intensity of the domain diffuse scattering. The structural expansion of the PTO/STO SL in the out-of-plane direction is apparent in the $Q_z$ dependence of the diffracted intensity of the SL Bragg reflection shown as a function of the optical intensity in Fig. 3(a). At absorbed intensities of more than 0.7 W/cm$^2$, the SL Bragg reflections splits and develops a new intensity maximum shifted to lower wavevector, indicating that a fraction of the volume of the SL is expanded. The diffracted intensity of the domain scattering in Fig. 3(b) decreases with increasing optical intensity and exhibits no optically induced shift along $Q_z$. There is no domain diffuse scattering around the shifted SL Bragg reflection. These changes in the domain scattering indicate that the nanodomain population remains only in an untransformed region of the SL. A similar effect is observed in the electric-field-induced transformation to the uniform polarization state in a similar PTO/STO SL [17]. Further evidence for the coexistence of nanodomain region and uniform-polarization region is obtained by comparing the integrated intensities of the unshifted fraction of the SL Bragg reflection and the domain diffuse scattering. The integrated intensities of the domain diffuse scattering and unshifted SL Bragg reflection have the same dependence on optical intensity, indicating that the remaining diffuse scattering arises from regions of untransformed SL.

The dependence of structural expansion and domain intensity on the optical intensity is shown in Figs. 4(a) and 4(b). Figure 4(a) shows the variation of the lattice parameter of the SL for optical intensities from 0 to 7.4 W/cm$^2$. Intensities below 1.3 W/cm$^2$ produce a negligible change in the lattice parameter. The uniform polarization state is favored is favored at high optical intensity, with a transition at a threshold intensity. The saturation of the optically induced lattice expansion at high intensities suggests that the expansion arises through screening of the depolarization field, which saturates as the field is completely compensated by charge carriers [23]. The photoinduced out-of-plane expansion reaches 0.9% at 7.4 W/cm$^2$. As shown in Fig 4(b), the intensity of the domain diffuse scattering also changes negligibly for optical intensities below 1.3 W/cm$^2$. Above the threshold intensity, the domain diffuse scattering intensity drops dramatically but does not completely disappear even at the intensity at which the lattice expansion saturates.

A temperature-dependent laboratory x-ray scattering study was conducted to evaluate the possibility that the reduction of the domain diffuse scattering and expansion of the SL arise from thermal, rather than optically driven, effects. The temperature dependence of the SL lattice parameter is shown in Fig. 4(c). In contrast to the dependence on the absorbed optical intensity, Fig. 4(c) shows that the lattice parameter decreases at elevated temperature, consistent with the heating of PTO-based thin films on STO [28]. Unlike the optical experiment, in which the STO substrate remains close to room temperature, both the SL and STO substrate were heated in the laboratory experiments. An elastic calculation converting the measured SL lattice parameters in the laboratory to the optically driven case, in which the lattice parameter of the substrate is constant and only the film is heated, also yields a contraction of the SL [37,38]. The linear decrease in domain diffuse scattering, shown in Fig. 4 (d), with increasing temperature arises because the domain diffuse scattering intensity is proportional to the square of polarization within the SL [39]. In the heating experiments, the domain intensity disappears at $T_C$, at the phase transition between ferroelectric and paraelectric states.

A LGD thermodynamic model was developed to provide insight into the origin of the lattice expansion and the threshold optical intensity for the reduction of the domain diffuse scattering (see Supplemental Materials) [40]. This approach extends a model of ferroelectric/dielectric SLs by Dawber *et al.* [8]. Comparing the free energies of the nanodomain and uniform polarization configurations shows that a uniform polarization is stable under conditions with high depolarization field screening. The screening of the depolarization field is described by a parameter $\theta$, which can range from 0 to 1. The uniform polarization state is energetically favorable for $\theta > 0.78$. For screening just above the critical value, the calculation predicts a lattice expansion of 0.32%, close to the 0.55% experimentally observed expansion at the 1.3 W/cm$^2$ threshold. The model predicts a saturation of the expansion at high values of $\theta$, which is also consistent with the experiment.

The LGD model exhibits an excellent match to the temperature-dependence of the domain scattering. The value of $T_C$ in the LGD model is 396 °C, in agreement with the experimental $T_C$ of 400 °C. The calculation also predicts that the nanodomain configuration is more stable than the uniform polarization state below $T_C$. The square of the calculated variation of the polarization with temperature has the dependence as the experimentally observed domain intensity, as in Fig. S2.

The timescales of the structural and domain pattern transformation provide insight into the mechanism of the optically induced transformation. The transient change in PTO/STO SL lattice parameter during and following illumination at absorbed intensity of 7.4 W/cm$^2$ is plotted in Fig. 5(a). The relaxation of the lattice expansion after illumination follows an approximately exponential time dependence with time constant $\tau$ = 2.3 s. The relaxation time is unexpectedly long in comparison with optically induced structural dynamics in uniform-polarization ferroelectric thin films. For example, Wen *et al.* observed a correlation between the optically induced structural response and nanosecond-scale carrier dynamics in $BiFeO_3$ [24]. In the present case, nanodomain patterns reemerge after the end of the illumination, instead, over several seconds as shown in Fig. 5(b), which has characteristic time 8.4 s.

Based on the experimental observations and the LGD model, we propose a transformation mechanism in which the depolarization field is screened by the trapping of excited charge carriers at defects. In this mechanism, trapped charges lead to a shift of the electron quasi-Fermi level and induce a population of mobile electrons, screening the depolarization field. Studies of above-bandgap illumination of ferroelectrics suggest that charge trapping occurs at surfaces, defects, and domain boundaries [41]. Theoretical studies indicate that oxygen vacancies can form easily in ferroelectric/dielectric superlattices [42]. The screening of the depolarization field by charges at oxygen vacancies or deep trapping centers has been theoretically predicted to enhance the polarization of ferroelectric-paraelectric heterostructures, an effect closely related to the lattice expansion we report [43]. Long time constants are also observed in the relaxation of trapped charges in illuminated ferroelectric thin films and capacitors [41,44]. Based on the value of $\theta$ at which the domain transformation is favored in LGD calculations, the charge density required to induce the domain transformation is $2\times10^{19}$ $cm^{-3}$ (see Supplemental Materials) [40]. The value of the required defect density is within the range of reported defect concentrations [28,44]. Polarization screening would be expected to lead to downward uniform polarization, based on the direction observed in other SLs [45].

In conclusion, we have shown that optical excitation can induce a transformation of the nanodomain pattern in a PTO/STO SL to a uniform polarization configuration, accompanied by an expansion of the lattice parameter. The existence of a threshold excitation for the domain transformation and the simultaneous structural expansion are consistent with predictions based on a depolarization field screening model. Based on the long timescales of transformation, the origin of the optically induced transformation appears to be linked to charge trapping in defects in the SL. The readily tunable structure of ferroelectric/dielectric SLs will allow optically induced effects to be incorporated into the design of new materials. By tuning the SL period or composition, for example, the magnitude of the polarization discontinuity at the interfaces and surfaces of SLs can be modified, which will allow control of the optically inducible strain. More generally, the relationship between the depolarization field and optically induced strain will provide the mechanism to probe the energetics of other exotic polarization configurations in complex oxide heterostructures.

X-ray scattering studies were supported by the U.S. DOE Office of Science Basic Energy Sciences through contract DE-FG02-04ER46147 (P.E.). Thin film synthesis efforts were supported by the U.S. NSF through grants DMR-1055413 and DMR-1334867 (M.D.). This research used resources of the Advanced Photon Source, a U.S. Department of Energy (DOE) Office of Science User Facility operated for the DOE Office of Science by Argonne National Laboratory under Contract No. DE-AC02-06CH11357.

FIG. 1. (a) Schematics and (b) x-ray reciprocal space maps for (i) room-temperature nanodomain configuration, (ii) optically induced uniform polarization state, and (iii) high-temperature paraelectric phase. Labels correspond to the reciprocal space locations of the reflections from the superlattice (SL) thin film, $SrRuO_3$ bottom electrode (SRO), and domain diffuse scattering (Domain). The dashed lines indicate the value of the out-plane-wavevector $Q_z$ at which the SL reflection appears at room temperature. (c) Experimental arrangement consisting of coincident focused optical pulses and focused x-ray nanobeam, illustrating the organization of nanodomains in the plane of the thin film and the composition of the SL.

FIG. 2. Maps of domain diffuse scattering intensity with (a) zero and (b) 1.3 W/cm$^2$ absorbed optical intensity. Diffracted intensities in planar sections of reciprocal space at $Q_x$ = 0 Å$^{-1}$ (top) and $Q_z$ = 3.1280 Å$^{-1}$ (bottom): (c) without optical excitation and (d) at an absorbed intensity of 1.3 W/cm$^2$.

FIG. 3. X-ray intensities as a function of $Q_z$ of (a) the PTO/STO (002) Bragg reflection and (b) (002) domain diffuse scattering at absorbed optical intensities from 0 to 7.4 W/cm$^2$.

FIG. 4. (a) Variation of the (a) PTO/STO lattice parameter and (b) intensity of the domain diffuse scattering as a function of absorbed optical intensity. Temperature dependence of (c) out-of-plane lattice parameter, and (d) integrated intensity of the domain diffuse scattering. Dashed lines in (a) and (c) indicate the SL lattice parameter in the absence of optical excitation at room temperature.

FIG. 5. Time dependence of (a) out-of-plane lattice parameter and (b) integrated domain diffuse scattering intensity at an absorbed intensity of 7.4 W/cm$^2$. The shaded area represents the duration over which the series of optical pulses illuminate the SL. The solid lines are an exponential relaxation fit to extract time constants.

Ahn *et al*., Figure 1.

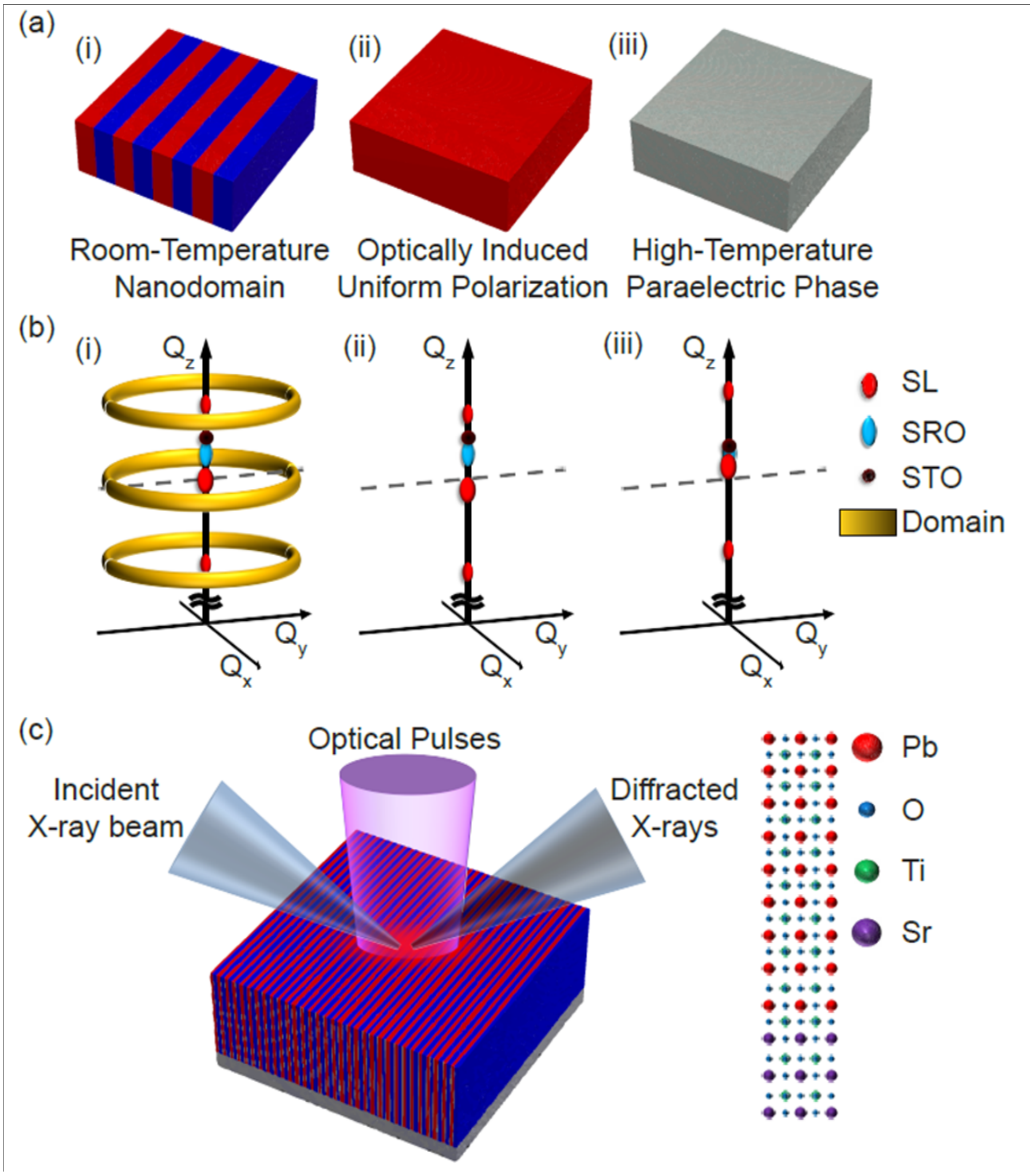

Ahn *et al*., Figure 2.

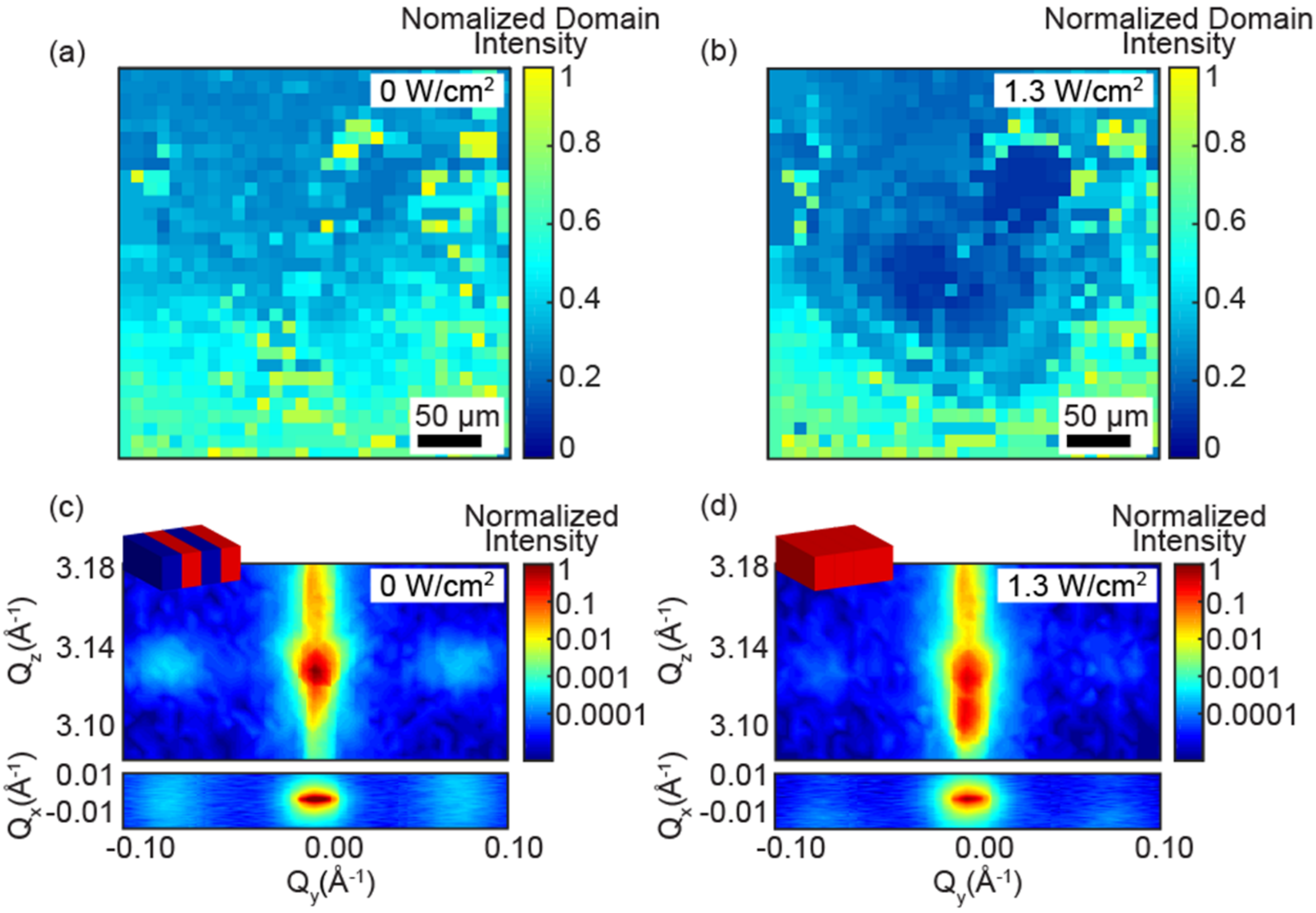

Ahn *et al*., Figure 3.

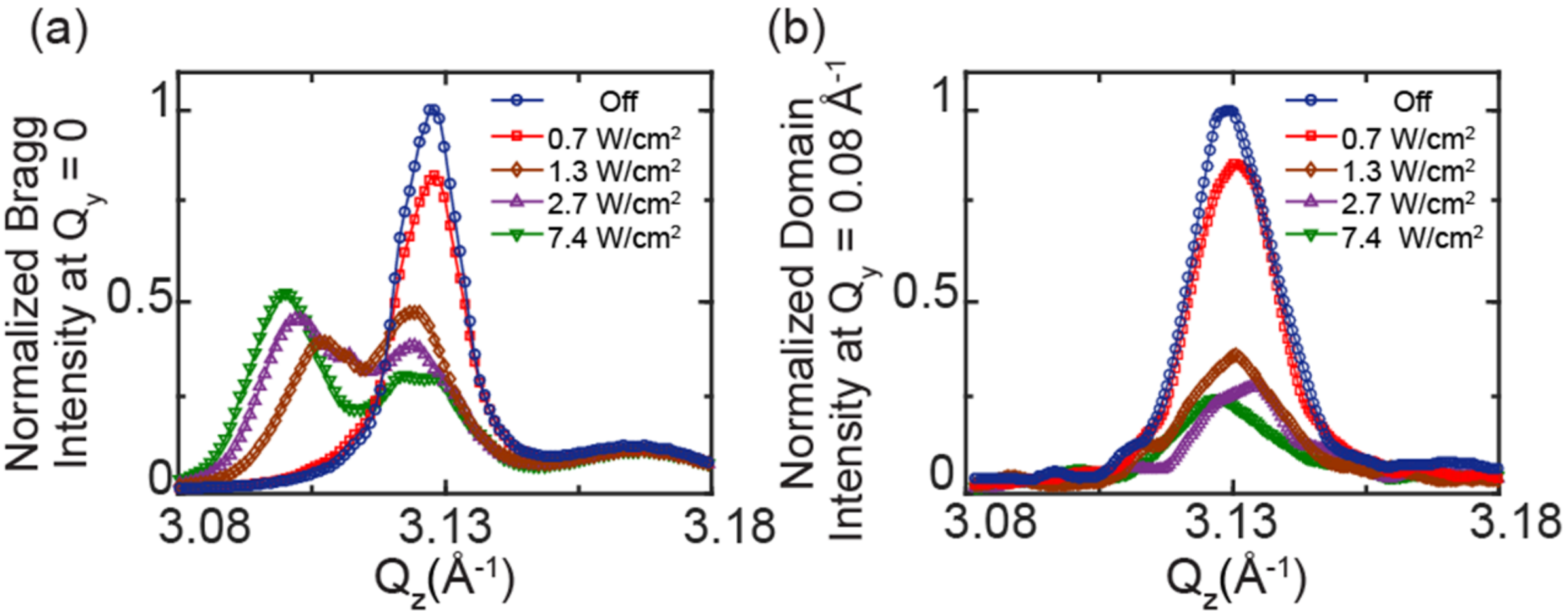

Ahn *et al*., Figure 4.

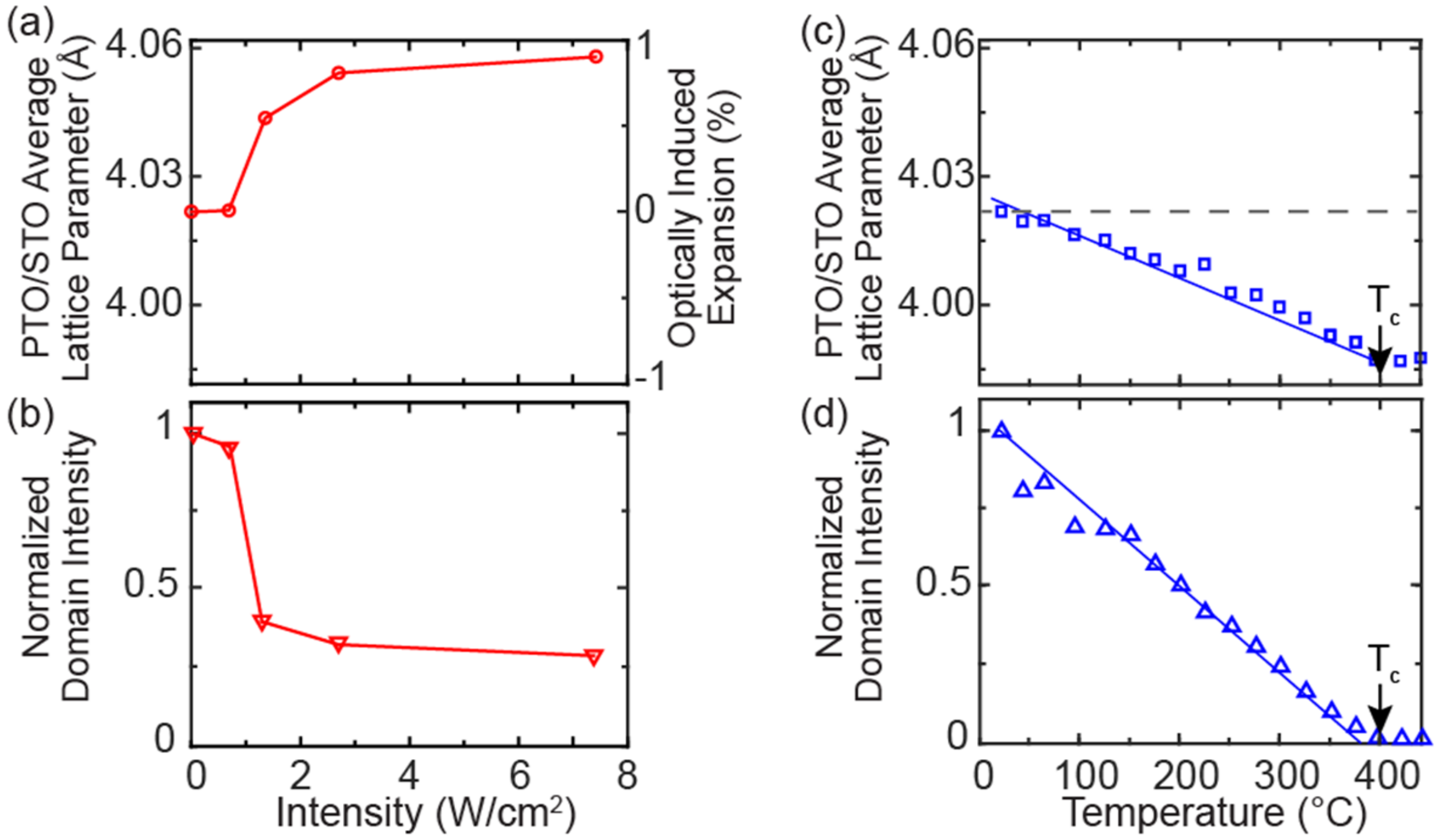

Ahn *et al*., Figure 5.

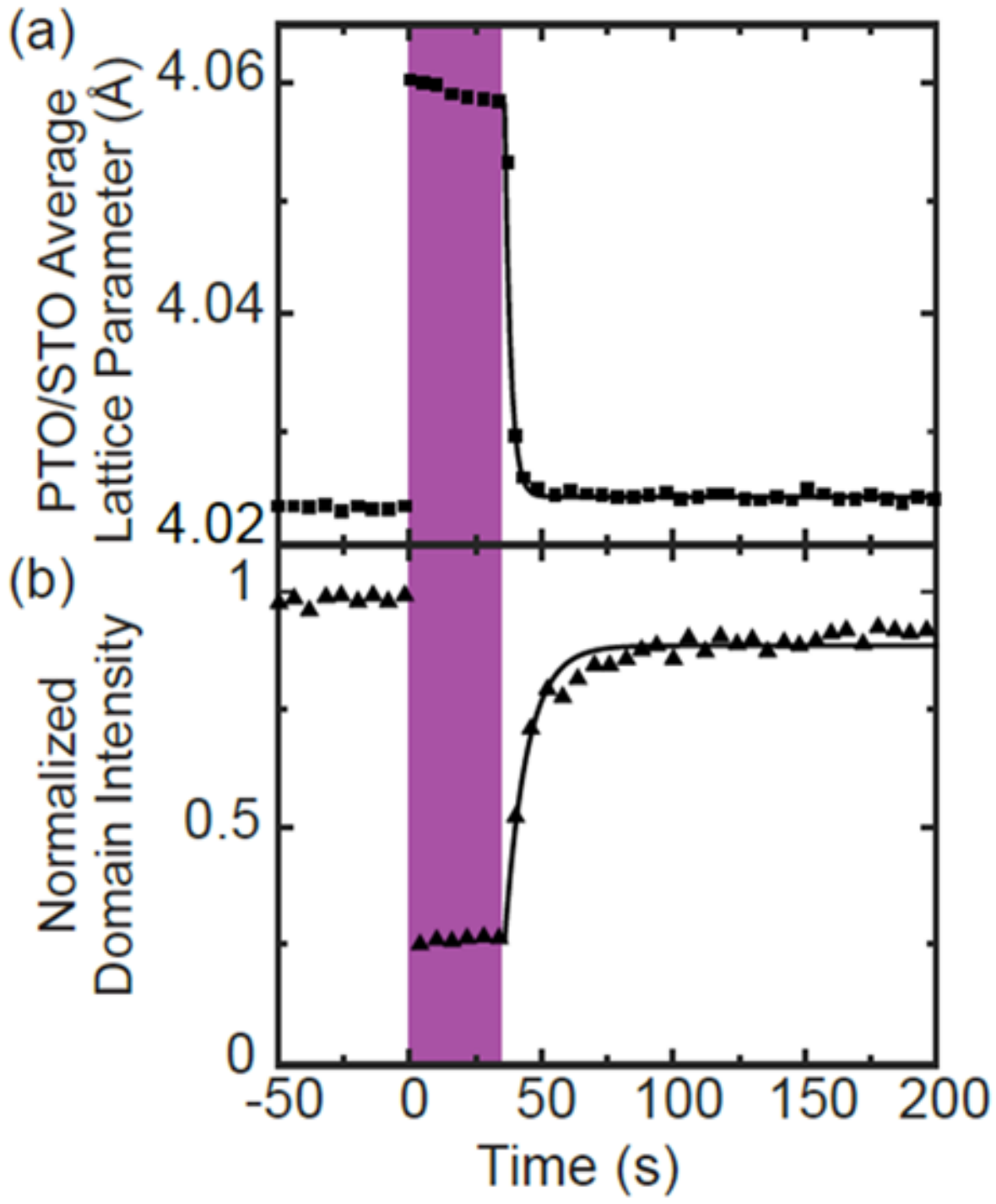

**Supplemental Material for**

# "Photoinduced Domain Pattern Transformation in Ferroelectric-Dielectric Superlattices"

Youngjun Ahn,[1] Joonkyu Park,[1] Anastasios Pateras,[1] Matthew B. Rich,[1] Qingteng Zhang,[1,*] Pice Chen,[1,*] Mohammed H. Yusuf,[2] Haidan Wen,[3] Matthew Dawber,[2] and Paul G. Evans[1,†]

[1]*Department of Materials Science and Engineering, University of Wisconsin-Madison, Madison, Wisconsin 53706, USA*

[2]*Department of Physics and Astronomy, Stony Brook University, Stony Brook, New York 11794, USA*

[3]*Advanced Photon Source, Argonne National Laboratory, Argonne, Illinois 60439, USA*

[*]Present address: Advanced Photon Source, Argonne National Laboratory, Argonne, Illinois 60439, USA

[†]pgevans@wisc.edu

## Landau-Ginzburg-Devonshire (LGD) thermodynamic calculations

A thermodynamic model based on LGD theory was developed to investigate the effect of the depolarization field $E_d$ on the domain configurations and lattice expansion. The free energy of the uniform polarization state can be described by $F = \Phi_P F_P + (1 - \Phi_P)F_S + E_{elec}$, where $\Phi_P$ is the volume fraction of PTO in the superlattice, $F_P$ and $F_S$ are the Helmholtz free energy densities of the PTO and STO layers, respectively, and $E_{elec}$ is the electrostatic energy density arising from the depolarization field. The formulation for the free energy density of PTO by Pertsev et al. [S1,S2] was used to take into account the mechanical boundary conditions: fixed compressive in-plane strain and zero out-of-plane stress. The in-plane polarization components, $P_1$ and $P_2$, were assumed to be zero, as predicted when the SL is grown on an STO substrate. The compressive misfit strain, $u_m$, was estimated to be -0.0164 from $u_m = (a_{substrate}-a_0)/a_{substrate}$ where $a_{substrate}$ is the in-plane lattice parameter of STO and $a_0$ is the in-plane cubic cell lattice constant of PTO film.

The Helmholtz free energy density of the $i^{th}$ component of the superlattice can be expressed in terms of out-of-plane polarization, $P$, as $F_i(P) = a_{3,i}P^2 + a_{33,i}P^4 + a_{111,i}P^6 + (c_{11,i}^2 + c_{11,i}^2 c_{12,i}^2 - 2c_{12,i}^2)u_{m,i}^2/c_{11,i}^2$ where $a_i$, $a_{ij}$, and $a_{ijk}$ are the Landau coefficients adjusted for the strain constraint and $c_{ij}$ is the elastic stiffness given in Refs. [S1] and [S2]. For the uniform polarization state, the electrostatic energy density is $E_{elec} = \varepsilon_0 \varepsilon_{SL} \mathcal{E}_d^2/2$ [S3]. The depolarization field $\mathcal{E}_d$ was calculated from $\nabla \cdot D = \rho_f$ to be $\mathcal{E}_d = -(1-\theta)P/(\varepsilon_0 \varepsilon_{SL})$ where $D$ is the electric displacement, $\rho_f$ is the photoexcited charge density, $\theta$ is the screening parameter and $\varepsilon_{SL}$ is the dielectric constant of the superlattice [S4]. For the nanodomain configuration, the electrostatic energy density for a simple periodic domain structure is $E_{elec} = 1.7(1-\theta)^2 P^2 \Lambda/[(4\pi h \varepsilon_0)(\varepsilon_a \varepsilon_c)^{1/2}]$ where $\Lambda$ is the period of the stripe domain pattern, $h$ is the film thickness, $\varepsilon_a$ and $\varepsilon_c$ are the dielectric constants of the film perpendicular and parallel to the polarization direction [S5]. The domain wall energy density were also considered to evaluate the energy for the polarization gradient near domain walls, $E_{wall} = 2\gamma_0 (P/P_0)^3/\Lambda$ where $\gamma_0$ is the 180° domain wall formation energy at 0 K and $P_0$ is the spontaneous polarization at 0 K. The domain wall formation energy of PTO is given in Ref. [S6] as 132 mJ/m$^2$ at 0 K, based on a first-principles calculation. Throughout this model, the polarization discontinuity at the interfaces of PTO and STO layers in the SL is assumed to be zero.

The equilibrium configurations were predicted by calculating the total free energy for the uniform polarization state and nanodomain configuration as a function of the screening parameter, $\theta$. The spontaneous polarization and relative dielectric constants at room temperature was obtained from $\partial F/\partial P = 0$ and $\partial^2 F/\partial P^2 = 1/(\varepsilon_0 \varepsilon_{SL})$, assuming that the susceptibility of the superlattice is much higher than unity. As shown in Fig. S1, the uniform polarization state is favored for $\theta > 0.78$, corresponding to a screened bound charge density of 0.31 C/m$^2$ assuming that there are no external contributions to screen the bound charge except for optically induced charge carriers. The out-of-plane strain, $S$, was estimated from the mechanical condition of zero out-of-plane stress ($\partial F/\partial S = 0$), to be $3.2\times10^{-3}$ of the lattice expansion at 0.78 of the screening parameter.

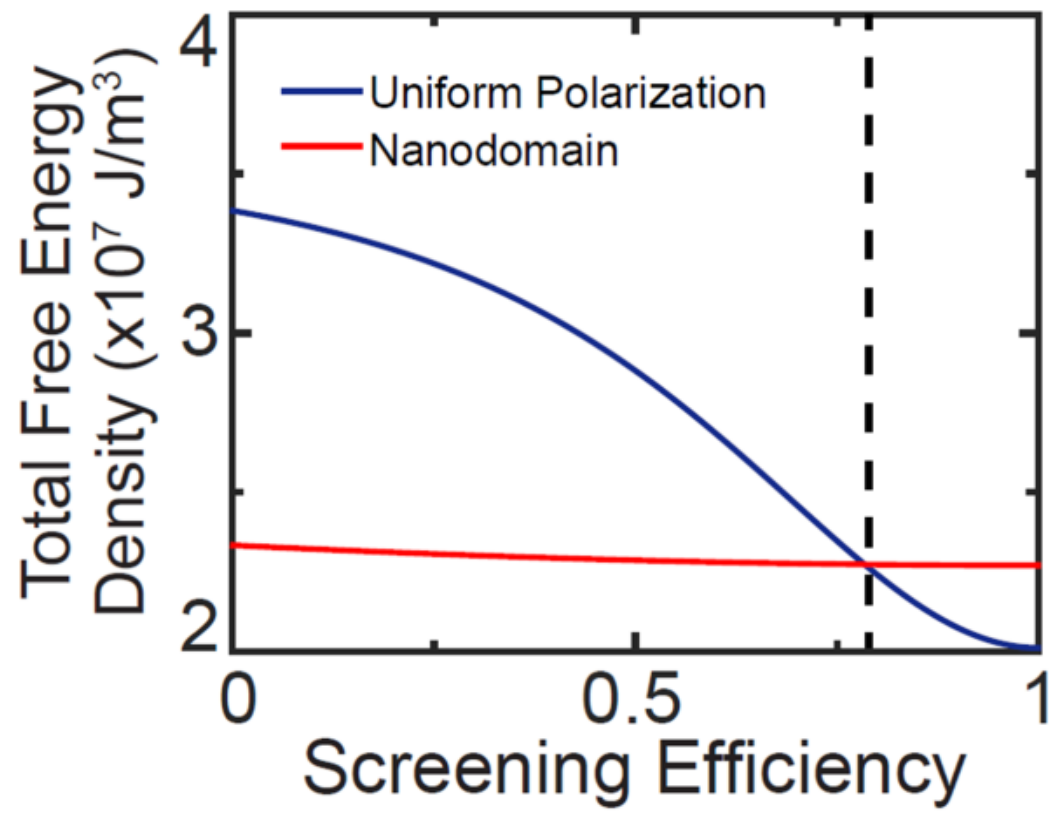

FIG. S1. Helmholtz free energy density of the uniform polarization state and nanodomain configuration as a function of the screening parameter, $\theta$. The dashed line indicates the transition between regimes in which the uniform polarization state and nanodomain configuration are stable.

The temperature-dependent control experiment was compared with the predicted values from the LGD model. The total free energy density for nanodomain and uniform polarization state with zero charge density is shown in Fig. S2(a). The experimental results show that the nanodomain configuration is stable for temperatures below the Curie temperature $T_C$, in agreement with the free energy density calculation. The predicted $T_C$ in Fig. S2(b) matches with the experimentally determined value $T_C$=400ºC. The domain diffuse scattering intensity from 180º nanodomain configuration is proportional to the square of polarization [S7]. Figure S2(c) shows normalized integrated domain intensity and normalized values of the square of the predicted polarization as a function of temperature.

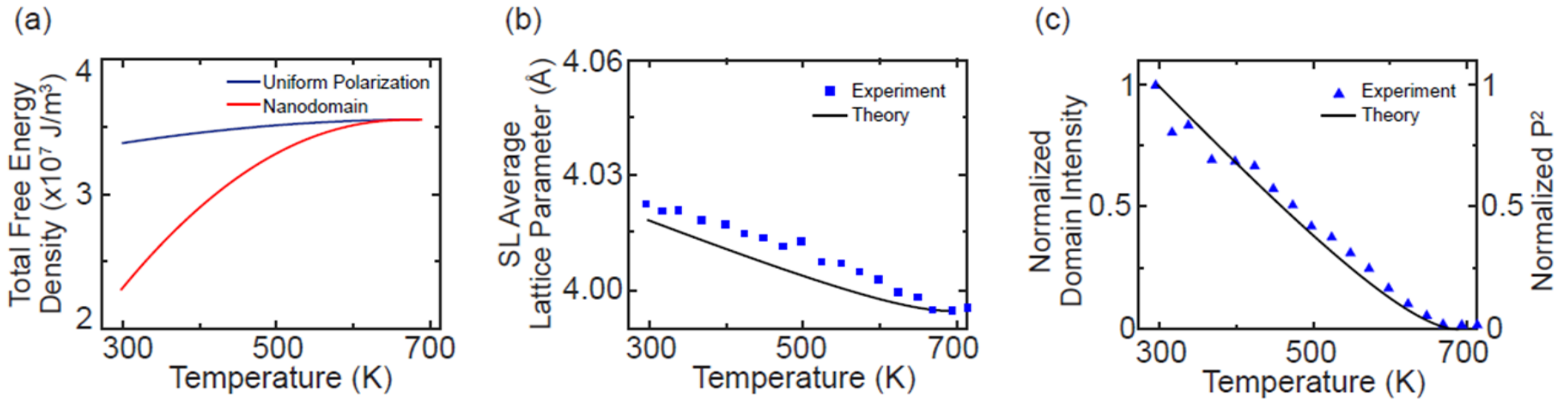

FIG. S2. (a) Free energy density of the uniform polarization state and nanodomain configurations. (b) PTO/STO SL average lattice parameter. (c) Normalized domain intensity and normalized values of the square of the estimated polarization from the LGD model as a function of temperature.

## References for Supplemental Material